\documentclass [reprint,amsmath,amssymb,aps,]{revtex4-1}
\usepackage{graphicx}
\usepackage{dcolumn}
\usepackage{bm}
\usepackage{mathrsfs}
\usepackage{braket}
\usepackage{xcolor}
\usepackage{listings}
\usepackage{setspace}
\begin{document}
\title{Waveguide-Integrated Mid-Infrared Plasmonics with High-Efficiency Coupling for Ultracompact Surface-Enhanced Infrared Absorption Spectroscopy}
\author{Daniel A. Mohr, Daehan Yoo, Che Chen, Mo Li, and Sang-Hyun Oh}
\email{address correspondence to sang@umn.edu}
\affiliation{Department of Electrical and Computer Engineering, University of Minnesota, Minneapolis, MN 55455, United States}
\date{\today}

\begin{abstract}
Waveguide-integrated plasmonics is a growing field with many innovative concepts and demonstrated devices in the visible and near-infrared. Here, we extend this body of work to the mid-infrared for the application of surface-enhanced infrared absorption (SEIRA), a spectroscopic method to probe molecular vibrations in small volumes and thin films. Built atop a silicon-on-insulator (SOI) waveguide platform, two key plasmonic structures useful for SEIRA are examined using computational modeling: gold nanorods and coaxial nanoapertures. We find resonance dips of 80\% in near diffraction-limited areas due to arrays of our structures and up to 40\% from a single resonator. Each of the structures are evaluated using the simulated SEIRA signal from poly(methyl methacrylate) and an octadecanethiol self-assembled monolayer. The platforms we present allow for a compact, on-chip SEIRA sensing system with highly efficient waveguide coupling in the mid-IR.
\end{abstract}

\maketitle

\section{Introduction}

Much work has previously been performed on integrating plasmonics\cite{ref1,ref2} with low-loss dielectric photonic integrated circuits (PICs) in an effort to miniaturize free-space optical components into a compact, chip-based system orders of magnitude smaller and at lower cost. Broad potential applications have already been explored, such as surface-enhanced Raman spectroscopy (SERS)\cite{ref3}, hybrid lasers\cite{ref4}, optical switching\cite{ref5}, directional waveguide coupling\cite{ref6}, plasmon-enhanced optical forces in waveguides\cite{ref7,ref8}, two-plasmon quantum interference\cite{ref9}, nanofocusing\cite{ref10,ref11}, integration with two-dimensional (2D) materials\cite{ref12,ref13,ref14}, and refractive index sensing\cite{ref15}. While most of this work is in the visible and near-infrared (NIR), the mid-infrared (MIR) has yet to be greatly explored. The MIR (typically 2-10 $\mu$m in wavelengths) is an important regime for biochemical spectroscopies thanks to the vast number of chemical resonances present that can give detailed information regarding molecular structure\cite{ref16,ref17,ref18,ref19}. The vibrational spectra of molecules are often measured with Fourier-Transform Infra-Red (FTIR) spectroscopy using broadband sources and free-space optics. However, with the advent of bright coherent MIR laser sources, MIR plasmonic antennas\cite{ref20}, the maturation of silicon photonics technology, and the growing interest in ultrasensitive chemical identification and diagnostics, there is tremendous potential for waveguide-integrated MIR systems. 

\begin{figure}
\includegraphics[width=3.3in]{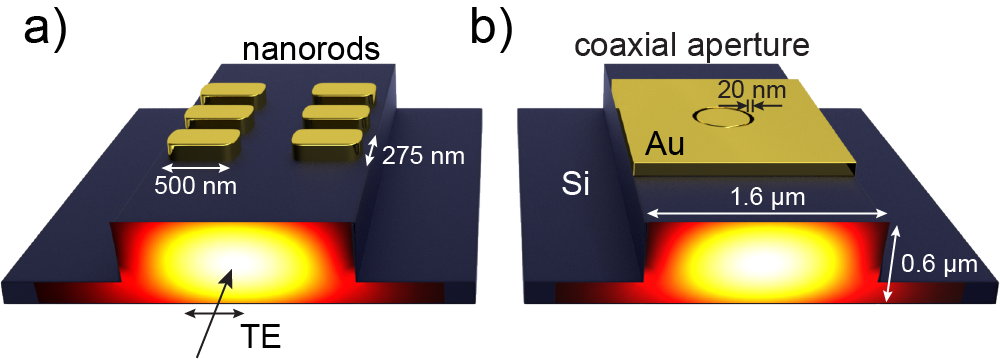}
\caption{\label{fig1} Schematics of the proposed devices for MIR waveguides integrated with plasmonics. (a) Nanorod pairs coupled together for high-efficiency coupling with minimal off-resonance scattering. (b) Coaxial nanoaperture embedded in a gold pad atop the waveguide for high coupling with a single resonator.}
\end{figure}

In this letter, we theoretically examine silicon waveguide-integrated plasmonics for surface-enhanced infrared absorption (SEIRA)\cite{ref21,ref22,ref23,ref24,ref25,ref26,ref27,ref28} spectroscopy using two common plasmonic resonator building blocks: nanorods and coaxial apertures (Figure \ref{fig1}). These structures have been studied in a free-space context for SEIRA, and have shown large performance improvements over standard infrared absorption techniques thanks to the field enhancement afforded by plasmonics, highly advantageous for thin films. Here, these same building blocks are arranged on a Si waveguide to perform obtain high coupling with a minimized footprint. The Si waveguide presented is designed based on a 600 nm silicon-on-insulator (SOI) wafer platform built for operation around 3 $\mu$m, a common location for resonances based on C-H bond stretching. To maintain single-mode operation, the width of the waveguide is limited to 1.6 $\mu$m, and a 200 nm thick support is used for releasing the waveguide from the oxide below. A schematic of the waveguide designed and the resonators proposed is available in Figure \ref{fig1}.

\section{Theory}
\begin{figure}
\includegraphics[width=3.3in]{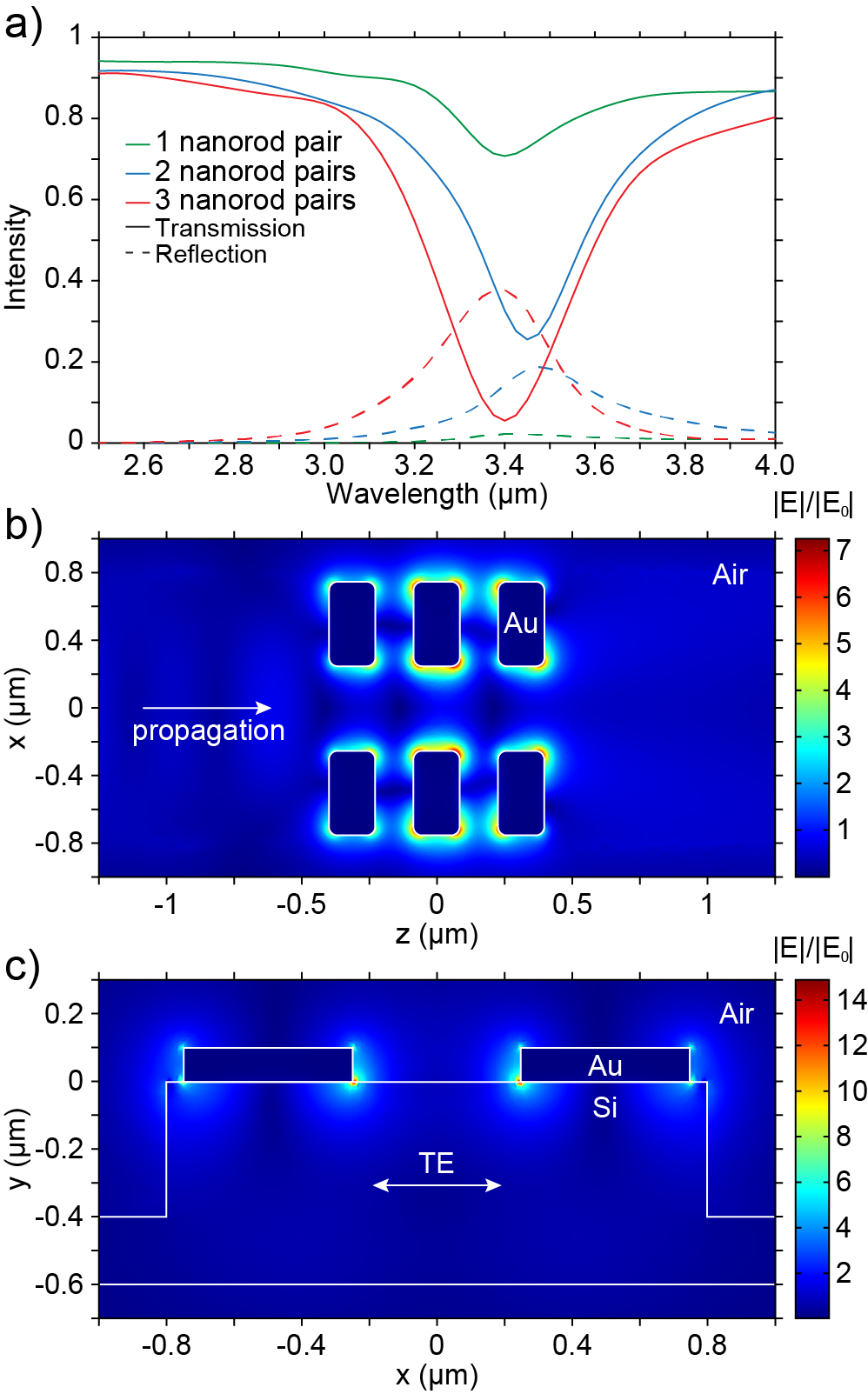}
\caption{\label{fig2} Nanorod-pair arrays integrated on a Si waveguide designed for the MIR. (a) Waveguide transmission of nanorod-pair arrays with different number of elements. Increasing the number of nanorod pairs increases the coupling obtained from the device. (b) Electric field distribution in a plane normal to the direction of waveguide propagation. (c) Same as (b), except in a plane taken at the top surface of the waveguide. Field enhancements are similar to those observed for far-field array devices (Figure \ref{figS1}). The nanorods have dimensions of 500 nm x 275 nm x 100 nm (LxWxH) with a radius of curvature of 50 nm for corners pictured in (b). The nanorod intra-pair spacing is 500 nm with a periodicity of 475 nm.}
\end{figure}

Metallic nanorods are a versatile building block for both waveguide-integrated plasmonics and resonant SEIRA. They are compact and can yield high field enhancements despite having such a simple geometry. While structures with few nanorods atop waveguides have already been fabricated and well-studied, we take this structure and tune it for use on MIR waveguides using dimensions similar to previous work in this area\cite{ref22,ref29}. Using these parameters combined with our waveguide design, we find that pairs of nanorods placed on the waveguide yield efficient structures with lower scattering outside of the resonant wavelength compared to arrays of single rods (Figures \ref{fig2}a and \ref{figS1}a). To further increase the resonance dip in waveguide transmission, multiple pairs are placed in close proximity to allow coupling (Figure \ref{fig2}a). We find that this design shows similar field enhancement to the far-field array version of this device (Figures \ref{fig2}b,c, and \ref{figS2}), while drastically reducing the excitation area from the spot-size of a typical FTIR light source ($\sim$100$\lambda^2$) to a diffraction-limited area ($\lambda^2/3$ when measured outside the waveguide). This allows for both the miniaturization to a chip-based device while also pushing the limit of detection closer to single-particle levels. In studying the device, we found that the antenna length, intra-antenna pair distance, and inter-antenna pair distance (i.e. period) are all critical in determining the resonance and coupling efficiency of the structure (Figures \ref{figS1}b-c). 

\begin{figure*}
\includegraphics[width=7in]{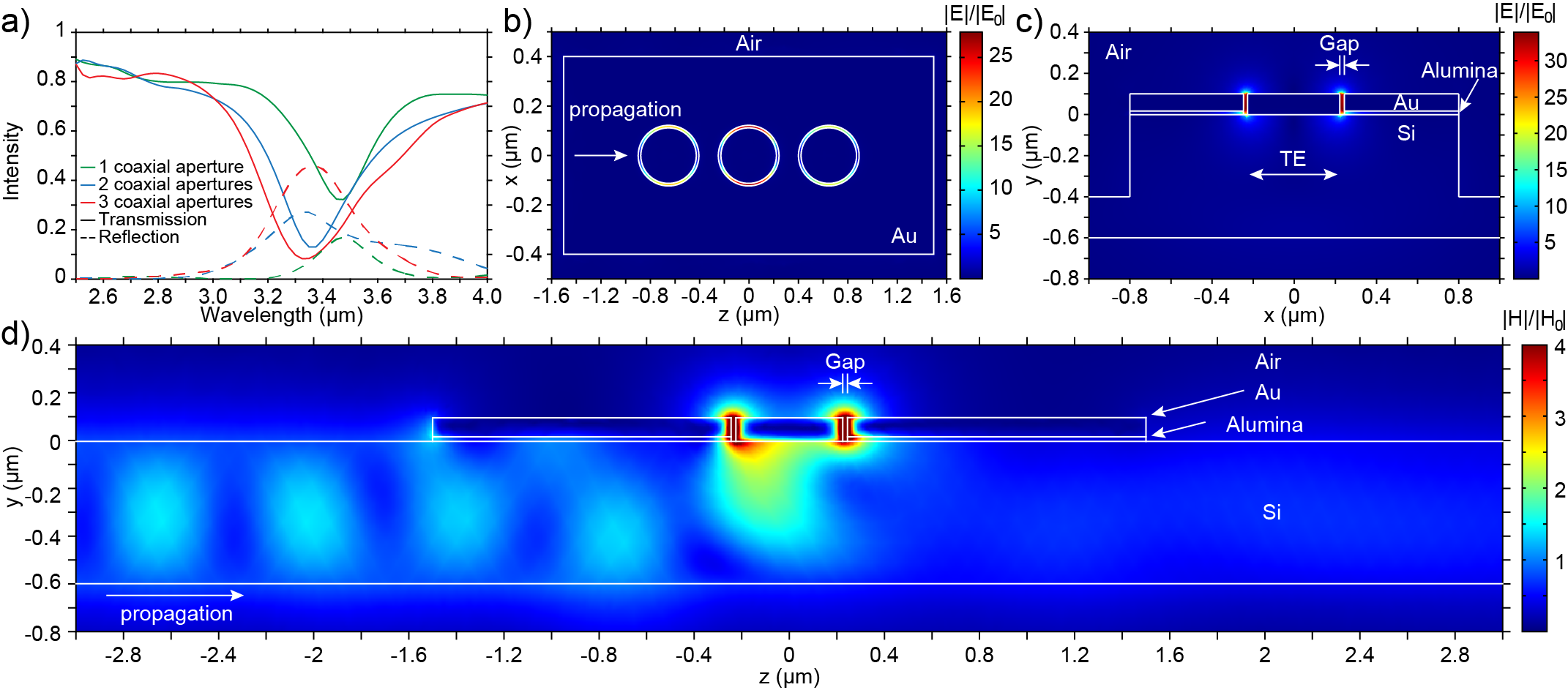}
\caption{\label{fig3} Coaxial apertures in a 3 $\mu$m long gold pad integrated on a MIR waveguide. (a) Comparison, between one, two, and three apertures placed in a serial array configuration atop the waveguide. (b) Electric field distribution taken at a plane halfway through the gold pad for the three-coaxial aperture device, demonstrating the highest coupling for the middle aperture. (c) Electric field distribution taken at a plane normal to the direction of waveguide propagation for the device with one aperture, demonstrating the relatively high and uniform field enhancement available in this device. (d) The magnetic field distribution of the device in (c), taken at a plane along the direction of propagation. The coaxial nanoapertures presented here are made in an 80 nm thick gold pad (3 $\mu$m long) with a 100 nm tall center conductor and based on the fabrication scheme presented by Yoo, et al\cite{ref30}, leaving a residual 20 nm spacer layer made of alumina beneath the gold pad. The inner radii of the apertures are 225 nm with a 20 nm gap and period of 650 nm between devices.}
\end{figure*}

A complementary structure we propose for waveguide-integrated plasmonics are annular apertures with very small gaps ($\sim$10s of nm), which are also promising for MIR SEIRA applications\cite{ref30,ref31}. These structures can support many different resonances\cite{ref30,ref32,ref33,ref34,ref35,ref36}, including a TEM mode\cite{ref37,ref38} that does not exhibit a cut-off, but the most convenient for excitation from a single TE-mode Si waveguide is the TE$_{11}$-like cutoff resonance. This cutoff resonance can be explained simply as a zeroth-order Fabry-Perot resonance. As the excitation wavelength increases to the cutoff wavelength, the real part of the propagation constant, $\beta$, decreases to zero, while the imaginary component increases, turning the mode from propagative to evanescent. Examining the transmission intensity for a Fabry-Perot cavity, $T=|t_1t_2|^2/|1-r_1r_2e^{i(2\beta d +\phi_1+\phi_2})|^2$ where $t_{1,2}$ are the transmission amplitude coefficients of the cavity ends, $r_{1,2}$ are the reflection amplitude coefficients with corresponding phases $\phi_{1,2}$, and $d$ is the thickness, we find that if the reflection phase is negative, it can cancel with $2\beta d$ and the Fabry-Perot condition is fulfilled, leading to resonant transmission and field enhancement. Since the propagation phase in the cavity is canceled upon reflection at the ends (and $2\beta d<2\pi$), the field profile is uniform, as seen in Figure \ref{fig3}c.  Alternatively, this zeroth-order Fabry-Perot resonance can be considered as an example of epsilon-near-zero (ENZ) phenomena\cite{ref39,ref40} and the high transmission resulting from ‘super-coupling’. As the radius of the aperture is increased, the resonance red shifts (toward longer wavelengths) and increases the coupling efficiency (Figure \ref{figS3}a). As the gap is decreased, the resonance also red-shifts (Figure \ref{figS3}b), but the strength of the resonance remains approximately the same, suggesting that mode-overlap area is more critical in coupling than the open-area of the aperture. 

Placing a single aperture on the waveguide with a 20 nm gap, we find coupling efficiencies of $\>$40\% around 3.5 $\mu$m (Figure \ref{fig3}a), much higher than those reported for individual nanorods\cite{ref41}. Placing multiple structures on the same gold pad increases the coupling as expected, with little dependence of the periodicity of the devices (Figure \ref{figS3}c). A fabrication method of these structures with open gaps available for the greatest SEIRA signal leverages atomic layer deposition (ALD) of sacrificial Al$_2$O$_3$ layers to define nanometer-wide gaps, a scheme called atomic layer lithography\cite{ref30,ref42,ref43}. In placing this structure on the waveguide, a metallic pad is needed to act as the cladding of the coaxial aperture, which we found to introduce relatively small scattering losses into the signal. Failure to properly design the cladding, however, can lead to quite significant pad resonances (Figure \ref{figS3}d).

\section{Results and Discussion}

\begin{figure}
\includegraphics[width=3.3in]{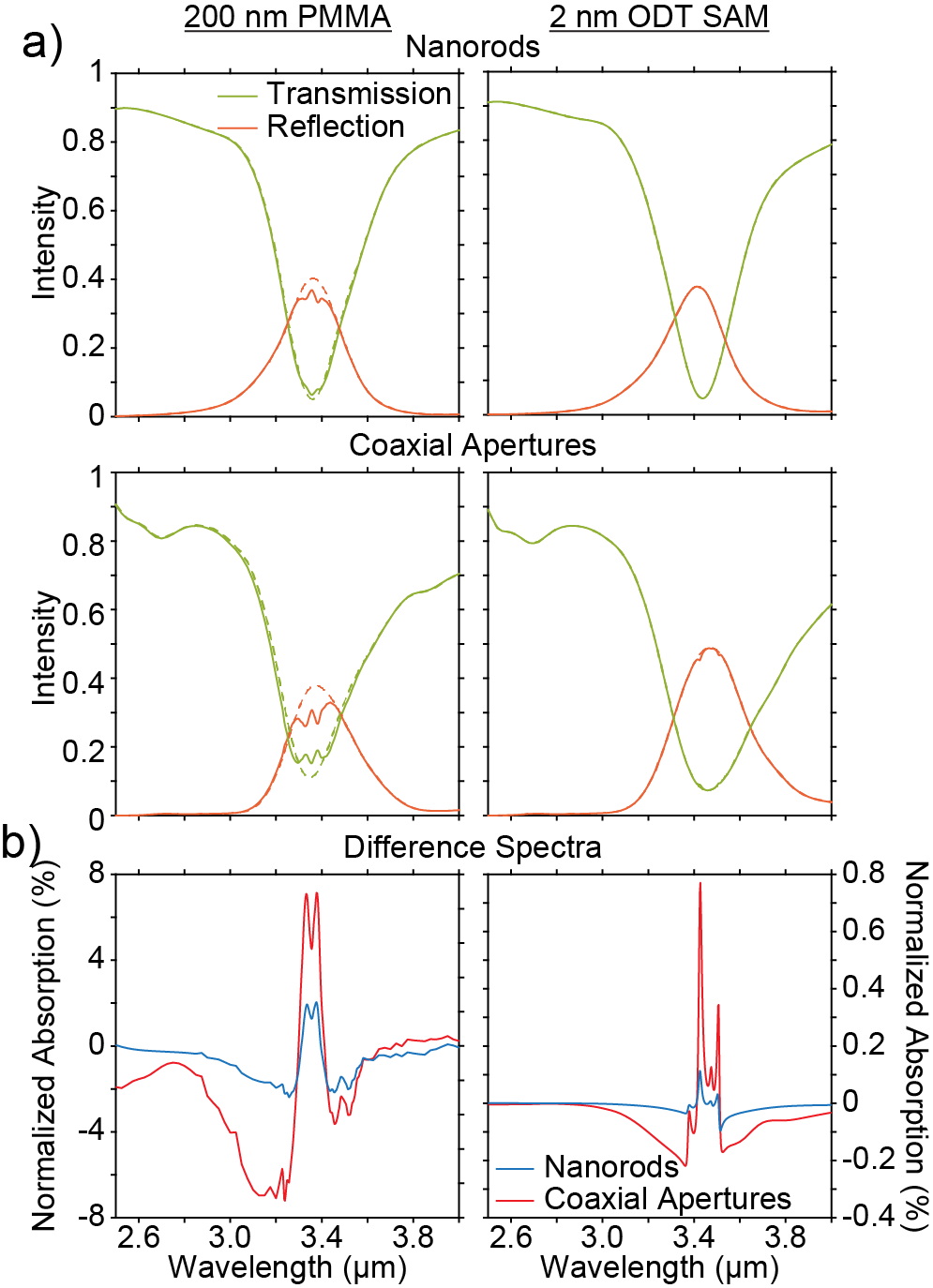}
\caption{\label{fig4} (a) Simulated SEIRA of PMMA and ODT coated on the two waveguide-integrated plasmonic structures (b) and corresponding difference signals calculated from the reflection spectra. The solid lines in (a) correspond to full PMMA and ODT dielectric functions, while the dashed lines correspond to lossless dielectric functions. The spectra in (b) were calculated by subtracting the lossy spectra from the lossless spectra and then normalizing to the lossless spectra.}
\end{figure}

To evaluate the SEIRA performance of each of these structures, we simulated the structures with both a uniform 200 nm layer of PMMA and a 2 nm octadecanethiol (ODT) self-assembled monolayer and examined the absorption of the C-H bond bending resonances around 3.4 $\mu$m. The 200 nm layer of PMMA is intended to probe the entire electric field distribution available for sensing, while the 2 nm layer yield information regarding the confinement of the field to the surface. As can be seen in Figure \ref{fig4}, both nanorods and coaxial nanoapertures yield clear absorption responses of 2\% and 7\%, respectively, for the 200 nm PMMA layer. With a 2 nm layer of ODT, the normalized absorption for a coaxial nanoaperture is over 0.7\%, while absorption in the nanorod case is much less, at just over 0.1\%. This is due to the relative resonance modes of each of the structures. For nanorods, a significant portion of the electric field distribution on resonance is contained inside the silicon, and therefore is not available for sensing (Figure \ref{fig2}b), while in etched coaxial nanoapertures, nearly the entire field distribution is available (Figure \ref{fig3}b). This explains some of the intuition of why coaxial nanoapertures in the 200 nm PMMA case yield higher absorption responses. When reducing the analyte thickness to only 2 nm, the metal nanorods yield greatly reduced signals due to the poor field confinement, while coaxial nanoapertures maintain higher absorption levels since the field confinement is essentially defined by the gap size, which is 20 nm in this case.

These findings suggest that coaxial nanoapertures generally exhibit higher performance, which is not surprising when taking into account the recent work of Huck, et al.\cite{ref44}, which describes why aperture-based structures are often superior to antennas. Essentially, the bulk of the nanorod signal is sourced from the ends of the device, while gaps generally have a more uniform, higher electric field distribution which increases sensitivity to analyte. The gap-defined field distribution in coaxial nanoapertures can be precisely controlled below 10 nm (e.g. by using atomic layer lithography). This allows one to both tailor the field distribution to the sensing analyte and increase the sensitivity of the device per analyte particle. With the devices considered in our work, we also found greater coupling per single resonator when comparing the two resonator designs, important for pushing to low limits of detection.

\begin{figure}
\includegraphics[width=3.3in]{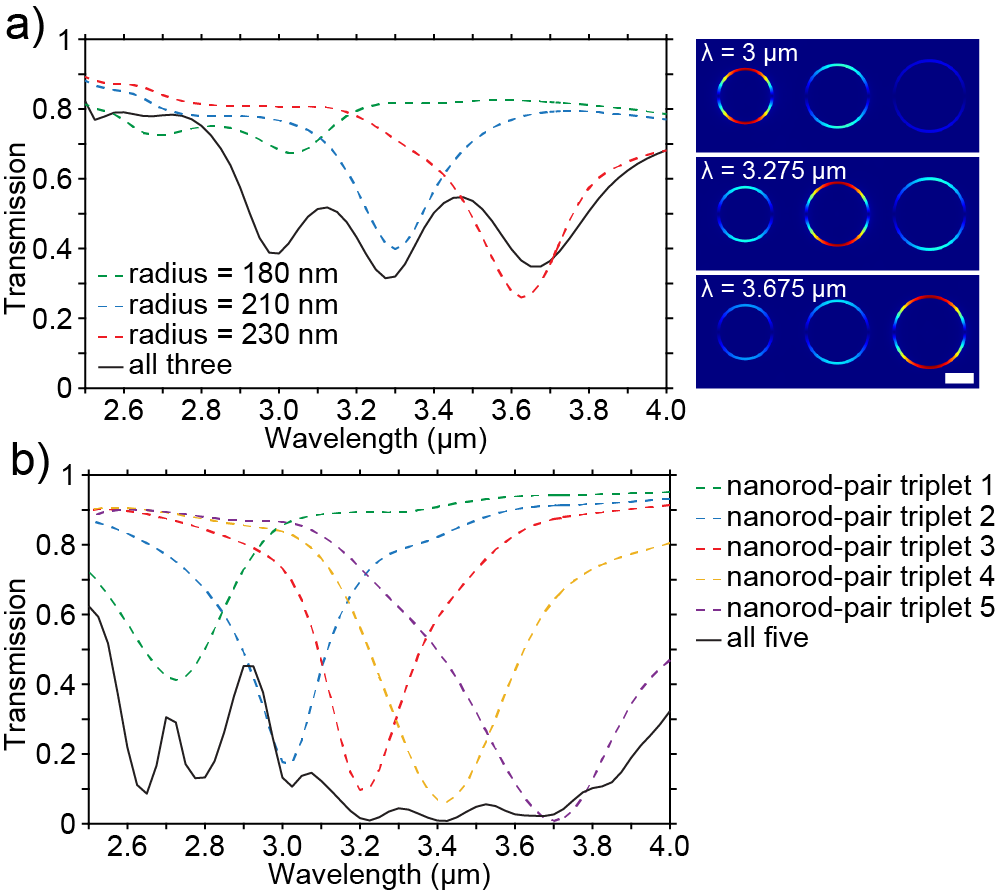}
\caption{\label{fig5} Plasmonic resonators with differing geometrical parameters can be fabricated atop a single MIR waveguide leading to broadband resonances, useful for probing a larger spectral area for SEIRA. (a) Resonance due to only three separate coaxial nanoapertures placed in a serial array made in one Au pad acting as the cladding. Associated field maps of the absorption dips are shown to the right. (b) Resonance dip from an array of five nanorod-pair triplets on the same waveguide. Scale bar in (a) is 200 nm.}
\end{figure}

Up to this point, structures with only one geometrical set of parameters have been examined. By placing multiple structures on the same waveguide with varying parameters, it is possible to create very broad resonances. As seen in Figure \ref{fig5}a, coaxial nanoapertures with varying radii can be combined into one small Au pad (3 $\mu$m long) to create a broadband resonance. Figure \ref{fig5}b contains the spectra simulated for five nanorod-pair triplets placed atop the same waveguide. Many other groups have been interested in creating plasmonic resonators for SEIRA that contain multiple resonances to obtain broad spectra of the analyte\cite{ref45}. While many of those same structures can still be used here (depending on their polarization sensitivity), by integrating devices serially on the waveguide, the same effect can be shown. While the work here has been targeted for SEIRA near 3 $\mu$m, the platform can easily be scaled to longer (or shorter) wavelength regimes, such as 6 $\mu$m, where a wide range of organic molecules have many strong vibrational modes. 

\section{Conclusion}

Waveguide-integrated plasmonics is a growing field with many promising aspects for use in the lab-on-a-chip and photonics communities. While visible/NIR sensing applications of both SERS and refractometric sensing have both been previously demonstrated, in this work we computationally push the devices to the MIR for the application of SEIRA. In comparing gold nanorods and coaxial nanoapertures using simulated PMMA and ODT layers, we find that ring-shaped coaxial nanoapertures, a gap-based device exhibiting super-coupling effects, yields superior performance to nanorods, but both can have $\sim$90\% absorption in a near diffraction-limited size. Furthermore, using a single coaxial nanoaperture, we find up to 40\% of absorption, pushing the bar towards ultrasensitive detections of low-number molecular analytes. While the devices are only theoretically studied here, the fabrication of these devices is achievable, which we plan to demonstrate in a future publication and evaluate real-world performance.

\section{Methods}Computational modeling was performed using COMSOL Multiphysics in the frequency domain. To decrease simulation time, only half of the geometry was simulated taking advantage of the plane of symmetry parallel to the waveguide while scattering boundary conditions were used perpendicular to the waveguide direction (except for the boundary of symmetry). Perfectly matched layers were used at the ends of the waveguide to increase the accuracy of the simulation. Reflection and transmission measurements were measured across a simulation plane before and after the plasmonic devices. Only the optical power in the fundamental TE modes were included in these calculations. Optical constants for silicon, gold, alumina, PMMA, and ODT were obtained from previous works\cite{ref46,ref47,ref48,ref49,ref50}.

\section{Acknowledgments}This work was supported primarily by the grant from the National Science Foundation (NSF ECCS No. 1610333 to D.Y. and S.-H.O.; NSF ECCS No. 1708768 to C.C. and M.L.); Seagate Technology through the University of Minnesota MINT consortium (D.A.M. and S.- H.O.). D.A.M. also acknowledges support from the NIH Biotechnology Training Grant (T32 GM008347) and Jackie. Computational modeling was carried out in part using resources provided by the University of Minnesota Supercomputing Institute.


\appendix
\section{Supporting Information}

\begin{figure*}[h]
\includegraphics[width=7in]{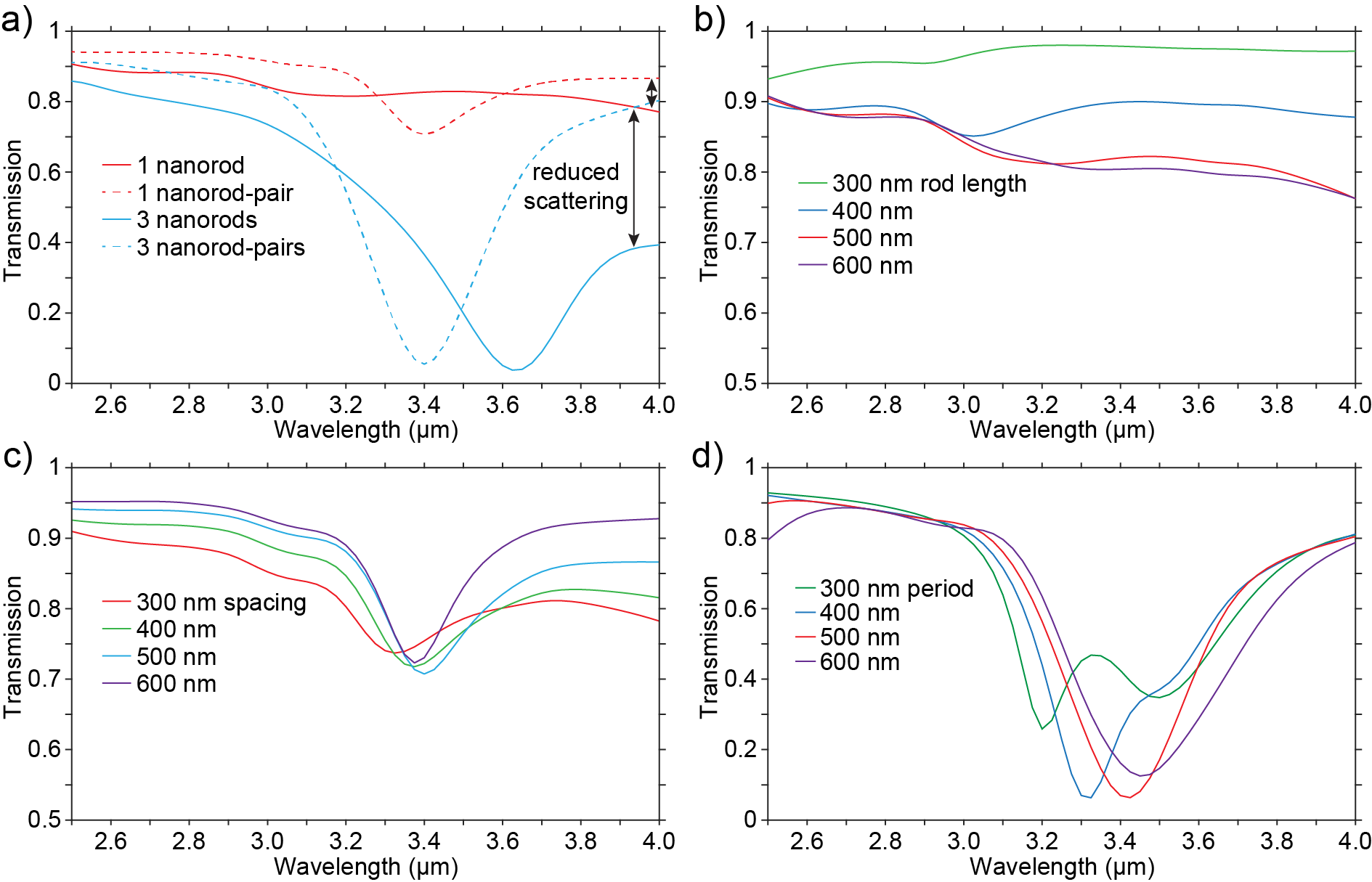}
\caption{\label{figS1} (a) Transmission spectra of individual nanorods and arrays of nanorods, along with nanorod-pair spectra. (b) Spectra of an individual nanorod on a waveguide with varied length. (c) Spectra of a single nanorod-pair with varying distance between the ends of the nanorods. (d) The transmission spectra of an array of three nanorod-pairs with varying periodicity.}
\end{figure*}

\begin{figure}[h]
\includegraphics[width=3.3in]{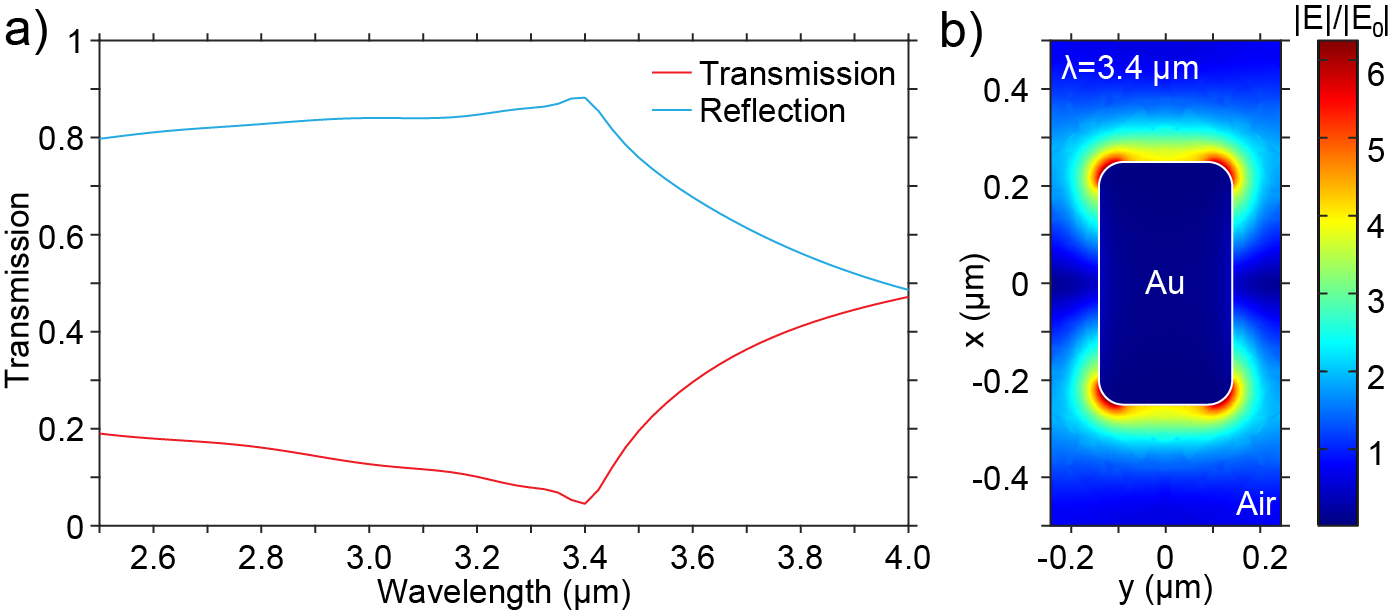}
\caption{\label{figS2}(a) Spectra of an array version of the nanorod structure used. Here, the rod length is 500 nm, the width is 275 nm, the period in the x-direction is 1 $\mu$m, and the period in the y-direction is 475 nm. (b) Corresponding electric field map exhibiting similar field enhancement as the waveguide-integrated device.}
\end{figure}

\begin{figure*}[h]
\includegraphics[width=7in]{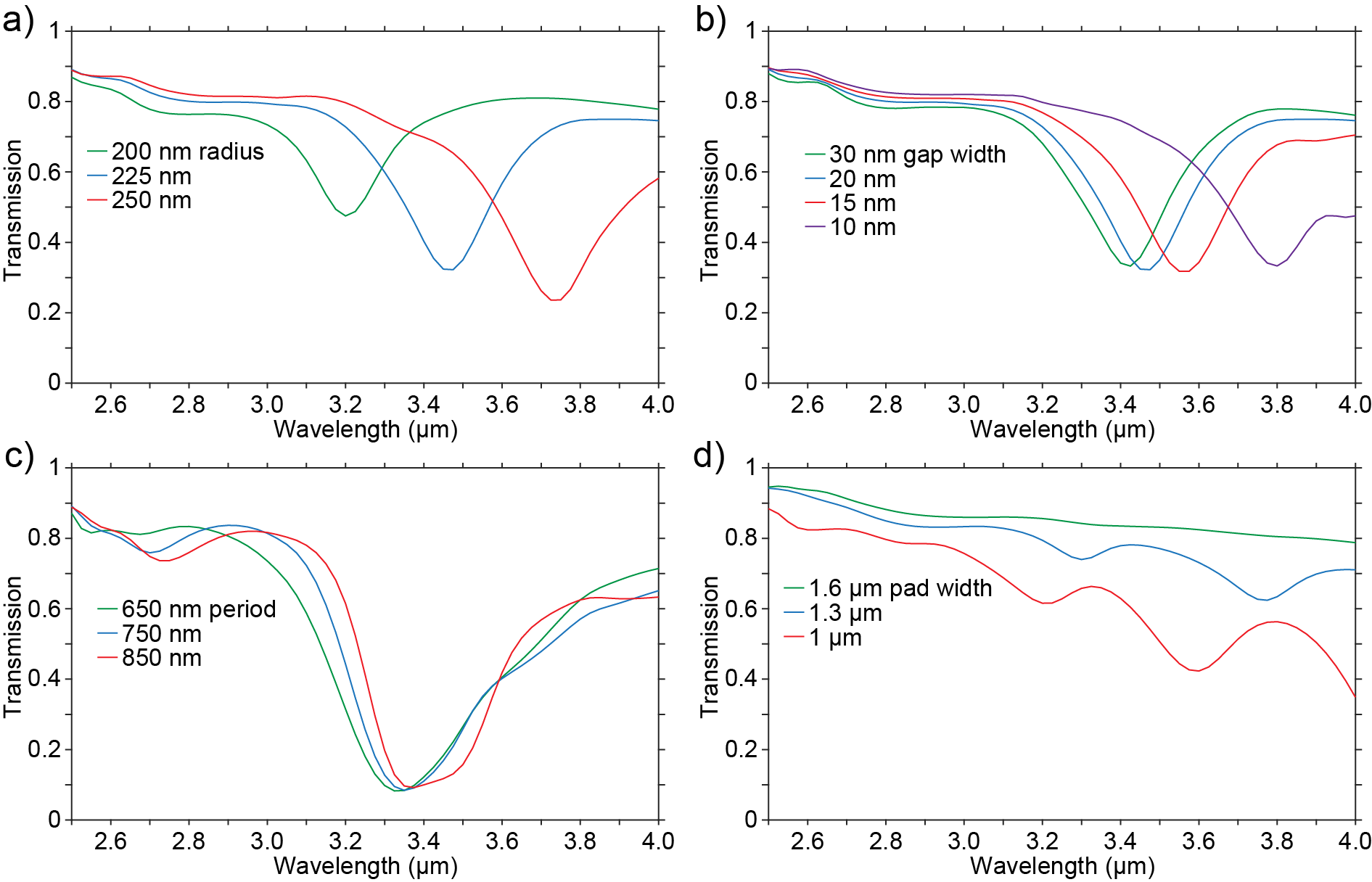}
\caption{\label{figS3}Spectra of individual coaxial nanoapertures with (a) three different radii and (b) four different gap widths. (c) Arrays of three coaxial nanoapertures with different periods. (d) Transmission spectra with only the gold pad present and no coaxial nanoaperture. The width of the gold pad is varied between 1 and 1.6 $\mu$m.}
\end{figure*}


\begin{thebibliography}{1}
\bibitem{ref1}Maier, S. A.; Kik, P. G.; Atwater, H. A.; Meltzer, S.; Harel, E.; Koel, B. E.; Requicha, A. A. G. Local Detection of Electromagnetic Energy Transport Below the Diffraction Limit in Metal Nanoparticle Plasmon Waveguides. Nature Mater. 2003, 2, 229-232.
\bibitem{ref2}Barnes, W. L.; Dereux, A.; Ebbesen, T. W. Surface Plasmon Subwavelength Optics. Nature 2003, 424, 824-830.
\bibitem{ref3}Peyskens, F.; Dhakal, A.; Van Dorpe, P.; Le Thomas, N.; Baets, R. Surface Enhanced Raman Spectroscopy Using a Single Mode Nanophotonic-Plasmonic Platform. ACS Photonics 2015, 3, 102-108.
\bibitem{ref4}Oulton, R. F.; Sorger, V. J.; Zentgraf, T.; Ma, R.-M.; Gladden, C.; Dai, L.; Bartal, G.; Zhang, X. Plasmon Lasers at Deep Subwavelength Scale. Nature 2009, 461, 629-632.
\bibitem{ref5}Haffner, C.; Heni, W.; Fedoryshyn, Y.; Niegemann, J.; Melikyan, A.; Elder, D. L.; Baeuerle, B.; Salamin, Y.; Josten, A.; Koch, U.; Hoessbacher, C.; Ducry, F.; Juchli, L.; Emboras, A.; Hillerkuss, D.; Kohl, M.; Dalton, L. R.; Hafner, C.; Leuthold, J. All-Plasmonic Mach–Zehnder Modulator Enabling Optical High-Speed Communication at the Microscale. Nature Photonics 2015, 9, 525-528.
\bibitem{ref6}Vercruysse, D.; Neutens, P.; Lagae, L.; Verellen, N.; Van Dorpe, P. Single Asymmetric Plasmonic Antenna as a Directional Coupler to a Dielectric Waveguide. ACS Photonics 2017, 4, 1398-1402.
\bibitem{ref7}Li, H.; Noh, J. W.; Chen, Y.; Li, M. Enhanced Optical Forces in Integrated Hybrid Plasmonic Waveguides. Opt. Express 2013, 21, 11839-11851.
\bibitem{ref8}Thijssen, R.; Verhagen, E.; Kippenberg, T. J.; Polman, A. Plasmon Nanomechanical Coupling for Nanoscale Transduction. Nano Lett. 2013, 13, 3293-3297.
\bibitem{ref9}Fakonas, J. S.; Lee, H.; Kelaita, Y. A.; Atwater, H. A. Two-Plasmon Quantum Interference. Nature Photonics 2014, 8, 317-320.
\bibitem{ref10}Choo, H.; Kim, M.-K.; Staffaroni, M.; Seok, T. J.; Bokor, J.; Cabrini, S.; Schuck, P. J.; Wu, M. C.; Yablonovitch, E. Nanofocusing in a Metal–Insulator–Metal Gap Plasmon Waveguide with a Three-Dimensional Linear Taper. Nature Photonics 2012, 6, 838-844.
\bibitem{ref11}Nielsen, M. P.; Lafone, L.; Rakovich, A.; Sidiropoulos, T. P. H.; Rahmani, M.; Maier, S. A.; Oulton, R. F. Adiabatic Nanofocusing in Hybrid Gap Plasmon Waveguides on the Silicon-on-Insulator Platform. Nano Lett. 2016, 16, 1410-1414.
\bibitem{ref12}Liu, M.; Yin, X.; Ulin-Avila, E.; Geng, B.; Zentgraf, T.; Ju, L.; Wang, F.; Zhang, X. A Graphene-Based Broadband Optical Modulator. Nature 2011, 474, 64-67.
\bibitem{ref13}Youngblood, N.; Chen, C.; Koester, S. J.; Li, M. Waveguide-Integrated Black Phosphorus Photodetector with High Responsivity and Low Dark Current. Nature Photonics 2015, 9, 247-252.
\bibitem{ref14}Chen, C.; Youngblood, N.; Peng, R.; Yoo, D.; Mohr, D. A.; Johnson, T. W.; Oh, S.-H.; Li, M. Three-Dimensional Integration of Black Phosphorus Photodetector with Silicon Photonics and Nanoplasmonics. Nano Lett. 2017, 17, 985-991.
\bibitem{ref15}F\'evrier, M.; Gogol, P.; Barbillon, G.; Aassime, A.; M\'egy, R.; Bartenlian, B.; Lourtioz, J.-M.; Dagens, B. Integration of Short Gold Nanoparticles Chain on SOI Waveguide Toward Compact Integrated Bio-Sensors. Opt. Express 2012, 20, 17402.
\bibitem{ref16}Maier, S. Plasmonics: Fundamentals and Applications; Springer: New York, NY, 2007.
\bibitem{ref17}Law, S.; Podolskiy, V.; Wasserman, D. Towards Nano-Scale Photonics with Micro-Scale Photons: the Opportunities and Challenges of Mid-Infrared Plasmonics. Nanophotonics 2013, 2, 103-130.
\bibitem{ref18}Rodrigo, D.; Limaj, O.; Janner, D.; Etezadi, D.; de Abajo, F. J. G.; Pruneri, V.; Altug, H. Mid-Infrared Plasmonic Biosensing with Graphene. Science 2015, 349, 165-168.
\bibitem{ref19}Low, T.; Avouris, P. Graphene Plasmonics for Terahertz to Mid-Infrared Applications. ACS Nano 2014, 8, 1086-1101.
\bibitem{ref20}Schnell, M.; Alonso-Gonzalez, P.; Arzubiaga, L.; Casanova, F.; Hueso, L. E.; Chuvilin, A.; Hillenbrand, R. Nanofocusing of Mid-Infrared Energy with Tapered Transmission Lines. Nature Photonics 2011, 5, 283-287.
\bibitem{ref21}Osawa, M. Surface-Enhanced Infrared Absorption. Topics Appl. Phys. 2001, 81, 163-187.
\bibitem{ref22}Adato, R.; Yanik, A. A.; Amsden, J. J.; Kaplan, D. L.; Omenetto, F. G.; Hong, M. K.; Erramilli, S.; Altug, H. Ultra-Sensitive Vibrational Spectroscopy of Protein Monolayers with Plasmonic Nanoantenna Arrays. Proc. Natl. Acad. Sci. USA 2009, 106, 19227-19232.
\bibitem{ref23}Wu, C.; Khanikaev, A. B.; Adato, R.; Arju, N.; Yanik, A. A.; Altug, H.; Shvets, G. Fano-Resonant Asymmetric Metamaterials for Ultrasensitive Spectroscopy and Identification of Molecular Monolayers. Nature Mater. 2012, 11, 69-75.
\bibitem{ref24}Brown, L. V.; Zhao, K.; King, N.; Sobhani, H.; Nordlander, P.; Halas, N. J. Surface-Enhanced Infrared Absorption Using Individual Cross Antennas Tailored to Chemical Moieties. J. Am. Chem. Soc. 2013, 135, 3688-3695.
\bibitem{ref25}Brown, L. V.; Yang, X.; Zhao, K.; Zheng, B. Y.; Nordlander, P.; Halas, N. J. Fan-Shaped Gold Nanoantennas Above Reflective Substrates for Surface-Enhanced Infrared Absorption (SEIRA). Nano Lett. 2015, 15, 1272-1280.
\bibitem{ref26}Huck, C.; Neubrech, F.; Vogt, J.; Toma, A.; Gerbert, D.; Katzmann, J.; H\''artling, T.; Pucci, A. Surface-Enhanced Infrared Spectroscopy Using Nanometer-Sized Gaps. ACS Nano 2014, 8, 4908-4914.
\bibitem{ref27}Chen, X.; Cirac\'i, C.; Smith, D. R.; Oh, S.-H. Nanogap-Enhanced Infrared Spectroscopy with Template-Stripped Wafer-Scale Arrays of Buried Plasmonic Cavities. Nano Lett. 2015, 15, 107-113.
\bibitem{ref28}Neubrech, F.; Huck, C.; Weber, K.; Pucci, A.; Giessen, H. Surface-Enhanced Infrared Spectroscopy Using Resonant Nanoantennas. Chem. Rev. 2017, 117, 5110-5145.
\bibitem{ref29}Chen, Y.; Lin, H.; Hu, J.; Li, M. Heterogeneously Integrated Silicon Photonics for the Mid-Infrared and Spectroscopic Sensing. ACS Nano 2014, 8, 6955-6961.
\bibitem{ref30}Yoo, D.; Nguyen, N.-C.; Mart\'in-Moreno, L.; Mohr, D. A.; Carretero-Palacios, S.; Shaver, J.; Peraire, J.; Ebbesen, T. W.; Oh, S.-H. High-Throughput Fabrication of Resonant Metamaterials with Ultrasmall Coaxial Apertures via Atomic Layer Lithography. Nano Lett. 2016, 16, 2040-2046.
\bibitem{ref31}Yoo, D.; Mohr, D. A.; Vidal-Codina, F.; John-Herpin, A.; Jo, M.; Kim, S.; Matson, J.; Caldwell, J. D.; Jeon, H.; Nguyen, N.-C.; Mart\'in-Moreno, L.; Peraire, J.; Altug, H.; Oh, S.-H. High-Contrast Infrared Absorption Spectroscopy via Mass-Produced Coaxial Zero-Mode Resonators with Sub-10 Nm Gaps. Nano Lett. 2018, 18, 1930-1936.
\bibitem{ref32}Baida, F. I.; Van Labeke, D. Light Transmission by Subwavelength Annular Aperture Arrays in Metallic Films. Opt. Commun. 2002, 209, 17-22.
\bibitem{ref33}Fan, W.; Zhang, S.; Minhas, B.; Malloy, K. J.; Brueck, S. R. J. Enhanced Infrared Transmission Through Subwavelength Coaxial Metallic Arrays. Phys. Rev. Lett. 2005, 94, 033902.
\bibitem{ref34}Orbons, S. M.; Roberts, A. Resonance and Extraordinary Transmission in Annular Aperture Arrays. Opt. Express 2006, 14, 12623-12628.
\bibitem{ref35}de Waele, R.; Burgos, S. P.; Polman, A.; Atwater, H. A. Plasmon Dispersion in Coaxial Waveguides From Single-Cavity Optical Transmission Measurements. Nano Lett. 2009, 9, 2832-2837.
\bibitem{ref36}Catrysse, P. B.; Fan, S. Understanding the Dispersion of Coaxial Plasmonic Structures Through a Connection with the Planar Metal-Insulator-Metal Geometry. Appl. Phys. Lett. 2009, 94, 231111.
\bibitem{ref37}Baida, F. I.; Belkhir, A.; Van Labeke, D.; Lamrous, O. Subwavelength Metallic Coaxial Waveguides in the Optical Range: Role of the Plasmonic Modes. Phys. Rev. B 2006, 74, 205419.
\bibitem{ref38}Li, D.; Gordon, R. Electromagnetic Transmission Resonances for a Single Annular Aperture in a Metal Plate. Phys. Rev. A 2010, 82, 041801(R).
\bibitem{ref39}Silveirinha, M.; Engheta, N. Theory of Supercoupling, Squeezing Wave Energy, and Field Confinement in Narrow Channels and Tight Bends Using $\epsilon$-Near-Zero Metamaterials. Phys. Rev. Lett. 2007, 76, 245109.
\bibitem{ref40}Argyropoulos, C.; Chen, P.-Y.; D'Aguanno, G.; Engheta, N.; Al\`u, A. Boosting Optical Nonlinearities in $\epsilon$-Near-Zero Plasmonic Channels. Phys. Rev. B 2012, 85, 045129.
\bibitem{ref41}Espinosa-Soria, A.; Mart\'inez, A.; Griol, A. Experimental Measurement of Plasmonic Nanostructures Embedded in Silicon Waveguide Gaps. Opt. Express 2016, 24, 9592-9601.
\bibitem{ref42}Im, H.; Bantz, K. C.; Lindquist, N. C.; Haynes, C. L.; Oh, S.-H. Vertically Oriented Sub-10-nm Plasmonic Nanogap Arrays. Nano Lett. 2010, 10, 2231-2236.
\bibitem{ref43}Chen, X.; Park, H.-R.; Pelton, M.; Piao, X.; Lindquist, N. C.; Im, H.; Kim, Y. J.; Ahn, J. S.; Ahn, K. J.; Park, N.; Kim, D.-S.; Oh, S.-H. Atomic Layer Lithography of Wafer-Scale Nanogap Arrays for Extreme Confinement of Electromagnetic Waves. Nature Commun. 2013, 4, 2361.
\bibitem{ref44}Huck, C.; Vogt, J.; Sendner, M.; Hengstler, D.; Neubrech, F.; Pucci, A. Plasmonic Enhancement of Infrared Vibrational Signals: Nanoslits Versus Nanorods. ACS Photonics 2015, 2, 1489-1497.
\bibitem{ref45}Limaj, O.; Etezadi, D.; Wittenberg, N. J.; Rodrigo, D.; Yoo, D.; Oh, S.-H.; Altug, H. Infrared Plasmonic Biosensor for Real-Time and Label-Free Monitoring of Lipid Membranes. Nano Lett. 2016, 16, 1502-1508.
\bibitem{ref46}Chandler-Horowitz, D.; Amirtharaj, P. M. High-Accuracy, Midinfrared (450 $\text{cm}^{-1}$ $\leq$ $\omega$ $\leq$ 4000 $\text{cm}^{-1}$) Refractive Index Values of Silicon. J. Appl. Phys. 2005, 97, 123526.
\bibitem{ref47}Olmon, R. L.; Slovick, B.; Johnson, T. W.; Shelton, D.; Oh, S.-H.; Boreman, G. D.; Raschke, M. B. Optical Dielectric Function of Gold. Phys. Rev. B 2012, 86, 235147.
\bibitem{ref48}Kischkat, J.; Peters, S.; Gruska, B.; Semtsiv, M.; Chashnikova, M.; Klinkm\''uller, M.; Fedosenko, O.; Machulik, S.; Aleksandrova, A.; Monastyrskyi, G.; Flores, Y.; Ted Masselink, W. Mid-Infrared Optical Properties of Thin Films of Aluminum Oxide, Titanium Dioxide, Silicon Dioxide, Aluminum Nitride, and Silicon Nitride. Appl. Opt. 2012, 51, 6789.
\bibitem{ref49}Graf, R. T.; Koenig, J. L.; Ishida, H. Optical Constant Determination of Thin Polymer Films in the Infrared:. Appl. Spectrosc. 1985, 39, 405-408.
\bibitem{ref50}Hu, Z. G.; Prunici, P.; Patzner, P.; Hess, P. Infrared Spectroscopic Ellipsometry of N-Alkylthiol (C$_{5}$-C$_{18}$) Self-Assembled Monolayers on Gold. J. Phys. Chem. B 2006, 110, 14824-14831.


\end{thebibliography}
\end{document}